\documentclass[aps,pra,preprint,titlepage,eqsecnum,floatfix,showkeys]{revtex4-2}

\usepackage{amsmath}
\usepackage{color}
\usepackage{graphicx}
\usepackage{hyperref}
\usepackage{multirow}

\newcommand{\bq}{\begin{eqnarray}}
\newcommand{\eq}{\end{eqnarray}}
\newcommand{\bqn}{\begin{eqnarray*}}
\newcommand{\eqn}{\end{eqnarray*}}
\newcommand{\bqs}{\begin{subequations}}
\newcommand{\eqs}{\end{subequations}}
\newcommand{\bw}{\begin{widetext}}
\newcommand{\ew}{\end{widetext}}

\newcommand{\vp}{{\varphi}}

\newcommand{\calo}{{\cal O}}

\newcommand{\red}[1]{{#1}}

\def\ra{\rightarrow}
\def\ds{d^s\!x}
\def\half{{\textstyle \frac{1}{2}}}

\def\ra{\rightarrow}

\def\k{\kappa}

\def\b{\begin{eqnarray*}}  %takes no eqn numbers
\def\e{\end{eqnarray*}}    %takes no eqn numbers
\def\bn{\begin{eqnarray}}  %takes eqn numbers
\def\en{\end{eqnarray}}    %takes eqn numbers

\begin{document}
%%%%%%%%%%%%%%%%%%%%%%%%%%%%%%%%%%%%%%%%%%%%%%%%%%%%%%%%%%%%%%%%%%%%%%%%%%%%%%
%%%%%%%%%%%%%%%%%%%%%%%%%%%%%%%%%%%%%%%%%%%%%%%%%%%%%%%%%%%%%%%%%%%%%%%%%%%%%%
%%%%%%%%%%%%%%%%%%%%%%%%%%%%%%%%%%%%%%%%%%%%%%%%%%%%%%%%%%%%%%%%%%%%%%%%%%%%%%
\title{Continuum limit of the Green function in scaled affine $\varphi^4_4$ quantum 
Euclidean covariant relativistic field theory}

\author{Riccardo Fantoni}
\email{riccardo.fantoni@scuola.istruzione.it}
\affiliation{Universit\`a di Trieste, Dipartimento di Fisica, strada
  Costiera 11, 34151 Grignano (Trieste), Italy}

\date{\today}

\begin{abstract}
We prove through path integral Monte Carlo computer experiments that the affine
quantization of the $\vp_4^4$ scaled Euclidean covariant relativistic scalar field 
theory is a valid quantum field theory with a well defined continuum limit of the 
one- and two-point-function. Affine quantization leads to a completely satisfactory 
quantization of field theories using situations that involve scaled behavior leading 
to an unexpected, $\hbar^2/\vp^2$ which arises only in the quantum aspects.
\end{abstract}

%\pacs{...}

\maketitle
%\tableofcontents
%%%%%%%%%%%%%%%%%%%%%%%%%%%%%%%%%%%%%%%%%%%%%%%%%%%%%%%%%%%%%%%%%%%%%%%%%%%%%%
\section{Introduction}
%%%%%%%%%%%%%%%%%%%%%%%%%%%%%%%%%%%%%%%%%%%%%%%%%%%%%%%%%%%%%%%%%%%%%%%%%%%%%%

It is well known that $\varphi^4_4$ quantum Euclidean covariant relativistic field 
theory when quantized through {\sl canonical} (Dirac \cite{Dirac}) {\sl quantization} 
(CQ) is \red{trivial} since its corresponding renormalized theory tends to a free 
theory in the continuum limit \cite{Freedman1982,Aizenman1981,Frohlich1982,Siefert2014,Wolff2014}.

Recently J. R. Klauder \cite{Klauder-FI,Klauder2000,Klauder2020c,Klauder2020,Klauder2020b} 
noticed that this difficulty can be overcome by using a different kind of quantization 
method, namely {\sl affine quantization} (AQ).

In a sequel of recent papers
\cite{Fantoni21c,Fantoni21e,Fantoni21f,Fantoni21h,Fantoni22a,Fantoni22b,Fantoni22c,
Fantoni22d,Fantoni23b,Fantoni23c,Fantoni23h} 
we proved, through path integral Monte Carlo (PIMC), that indeed affine quantization is 
able to make the $\varphi^4_4$ theory non-trivial. A crucial point left unanswered in 
these papers was the validity of the continuum limit at the level of the one- and 
two-point-functions. 

The aim of the present work is to show that as we approach the continuum on the 
computer, the one- and two-point-function converge to well defined results. In other 
words we prove the validity of the {\sl continuum limit} for the field theory quantized 
through affine quantization.  

Our \red{result} could \red{become important} in the physics of the standard model where the 
long-standing problem of the triviality of canonical quantum $\varphi^4$ theory is 
crucial for particle physics \red{since} it undermines the Higgs mechanism. It is also 
very important for progresses in quantum gravity where the role of the field is played
by the metric tensor which must be positive definite \cite{Fantoni23b}.
%%%%%%%%%%%%%%%%%%%%%%%%%%%%%%%%%%%%%%%%%%%%%%%%%%%%%%%%%%%%%%%%%%%%%%%%%%%%%%
\section{Field theory formulation}
%%%%%%%%%%%%%%%%%%%%%%%%%%%%%%%%%%%%%%%%%%%%%%%%%%%%%%%%%%%%%%%%%%%%%%%%%%%%%%

For a scalar field, $\vp$, with spacial degrees of freedom $x=(x_1,x_2,\ldots,x_s)$
and canonical momentum $\pi(x)$, the classical affine 
variables are $\k(x)\equiv \pi(x)\,\vp(x)$ and $\vp(x)\neq 0$. The reason we insist that 
$\vp(x)\neq0$ is because if $\vp(x)=0$ then $\k(x)=0$ whatever is $\pi(x)$.

We then introduce the classical Hamiltonian expressed in affine variables. This leads us 
to
\bn \label{eq:affine-H}
{\cal H}(\k,\vp) =\int\{\half[\k(x)^2\,\vp(x)^{-2}+(\nabla\vp(x))^2+m^2\,\vp(x)^2]+g\,\vp(x)^r\}\;\ds, 
\en
where $r$ is a positive, even, integer and $g\geq 0$ is the bare coupling constant such 
that for $g\to 0$ we fall into the free field theory. With these variables we do not let 
$\vp(x)=\infty$ otherwise $\vp(x)^{-2}=0$ which is not fair to $\k(x)$ and, as we already 
observed, we must forbid also $\vp(x)=0$ which would admit $\vp(x)^{-2}=\infty$ giving 
again an undetermined kinetic term. Therefore the AQ bounds $0<\vp(x)<\infty$ {\it 
forbid any \red{triviality}} 
\cite{Fantoni21c,Fantoni21e,Fantoni21f,Fantoni21h,Fantoni22a,Fantoni22b,Fantoni22c,
Fantoni22d,Fantoni23b,Fantoni23c,Fantoni23h}
which is otherwise possible for CQ
\cite{Freedman1982,Aizenman1981,Frohlich1982,Siefert2014}.

The quantum affine operators are the scalar field $\hat{\vp}(x)=\vp(x)$ and the 
{\it dilation} operator 
$\hat{\k}(x)=[\hat{\vp}(x)\hat{\pi}(x)+\hat{\pi}(x)\hat{\vp}(x)]/2$ 
where the momentum operator is $\hat{\pi}(x)=-i\hbar\delta/\delta\vp(x)$. Accordingly for 
the self adjoint kinetic term 
$\hat{\k}(x)\hat{\vp}(x)^{-2}\hat{\k}(x)=\hat{\pi}(x)^2+(3/4)\hbar\delta(0)^{2s}\vp(x)^{-2}$ 
(note that the factor $3/4$ that holds for $\vp>0$ should be replaced by a factor $2$ if 
$|\vp|>0$ \cite{Fantoni22b}) and one finds for the quantum Hamiltonian operator
\bn \label{eq:HO}
\hat{H}(\hat{\k},\hat{\vp}) =\int\left\{\half[\hat{\pi}(x)^2+(\nabla\vp(x))^2+m^2\,\vp(x)^2]+g\,\vp(x)^r +{\textstyle\frac{3}{8}}\hbar^2\frac{\delta(0)^{2s}}{\vp(x)^2}\right\}\;\ds.
\en

The affine action is found adding time, $x_0=ct$, where $c$ is the speed of light 
constant and $t$ is the Euclidean imaginary time, so that ${\cal S}=\int_0^\beta H\,dx_0$, 
with $H$ the semi-classical Hamiltonian corresponding to the one of Eq. (\ref{eq:HO}), will 
then read (see the appendix)
\bn \label{eq:action}
{\cal S}[\vp]=\int_0^\beta dx_0\,\int_{L^s}d^sx\,\left\{\half\left[\sum_{\mu=0}^s\left(\frac{\partial\vp(x)}{\partial x_\mu}\right)^2+m^2\,\vp(x)^2\right]+g\,\vp(x)^r+{\textstyle\frac{3}{8}}\hbar\frac{\delta(0)^{2s}}{\vp(x)^2}\right\}, 
\en
where with an abuse of notation we here use $x$ for $(x_0,x_1,x_2,\ldots,x_s)$ and 
$\beta=1/k_BT$, with $k_B$ the Boltzmann's constant and $T$ the absolute temperature. 
In this work we will set $\beta=L$. 

The vacuum expectation value of an observable ${\cal O}[\vp]$ will then be given by the 
following expression
\bn \label{eq:EV}
\langle{\cal O}\rangle=\frac{\int{\cal O}[\vp]\exp(-{\cal S}[\vp])\;{\cal D}\vp(x)}{\int\exp(-{\cal S}[\vp])\;{\cal D}\vp(x)},
\en
where the functional integrals will be calculated on a lattice using the PIMC method as
 explained later on.

The theory considers a real scalar field $\vp$ taking the value $\vp(x)$ on each site 
$x$ of a periodic $n$-dimensional lattice, with $n=s+1$ space-time dimensions, of lattice 
spacing $a$, the ultraviolet cutoff, spacial periodicity $L=Na$ and temporal 
periodicity $\beta=Na$. The field path is a closed loop on an $n$-dimensional closed 
surface of an $(n+1)$-dimensional $\beta$-periodic cylinder of radius $L$: an 
$(n+1)$-dimensional torus.
We used a lattice formulation of the AQ field theory of Eq. (\ref{eq:action}) (also 
studied in Eq. (8) of \cite{Fantoni21c}) using additionally the scaling 
$\vp\ra a^{-s/2}\vp$ and $g\ra a^{s(r-2)/2} g$ which is necessary to eliminate the 
Dirac delta factor $\delta(0)=a^{-1}$ divergent in the continuum limit $a\to 0$. The 
affine action for the field (in the {\it primitive approximation} \cite{Ceperley1995}) is 
then approximated by
\bn \label{eq:scaled-affine-action}
\frac{S[\vp]}{a}=\half\left\{\sum_{x,\mu}a^{-2}[\vp(x)-\vp(x+e_\mu)]^2 
+m^2\sum_{x}\vp(x)^2\right\}+\sum_{x}\left[g\,\vp(x)^r+{\textstyle\frac{3}{8}}{\displaystyle\frac{\hbar^2}{\vp(x)^2}}\right],
\en
where $e_\mu$ is a vector of length $a$ in the $+\mu$ direction with 
$\mu=0,1,2,\ldots,s$. We will have ${\cal S}\approx S$.

In this work we are interested in reaching the continuum limit by taking $Na$ fixed 
and letting $N\to\infty$ at fixed volume $L^s$.

We performed a PIMC \cite{Metropolis,Kalos-Whitlock,Ceperley1995,Fantoni12d} 
calculation for the AQ field theory described by the action of Eq. 
(\ref{eq:scaled-affine-action}) in natural Planck units $c=\hbar=k_B=1$. Specifically
we studied the $s=3$ and $r=4$ case. We calculated the renormalized coupling 
constant $g_R$ and mass $m_R$ defined in Eqs. (11) and (13) of \cite{Fantoni21c} 
respectively, measuring them in the PIMC through vacuum expectation values 
like in Eq. (\ref{eq:EV}).

In particular:
\bn
m_R^2=\frac{p_0^2\langle|\tilde{\vp}(p_0)|^2\rangle}{\langle\tilde{\vp}(0)^2\rangle-\langle|\tilde{\vp}(p_0)|^2\rangle},
\en 
and at zero momentum
\bn
g_R=\frac{3\langle\tilde{\vp}(0)^2\rangle^2-\langle\tilde{\vp}(0)^4\rangle}{\langle\tilde{\vp}(0)^2\rangle^2},
\en
where $\tilde{\vp}(p)=\int d^nx\;e^{ip\cdot x}\vp(x)$ is the Fourier transform of the field 
and we choose the 4-momentum $p_0$ with one spacial component equal to $2\pi/Na$ and all 
other components equal to zero.

We also calulated the 
one-, two-point-, and two-point-connected-function, respectively given by
\bq
V     &=&\sum_x\langle\varphi(x)\rangle/N^n,\\
D(z)  &=&\sum_x\langle\varphi(x)\varphi(x+z)\rangle/N^n,\\
D_c(z)&=&\sum_x(\langle\varphi(x)\varphi(x+z)\rangle-\langle\varphi(x)\rangle^2)/N^n
=D(z)-V^2.
\eq
By construction, these are periodic functions, $D(z)=D(z+L)$, of period $L$. Moreover, 
since the action ${\cal S}$ contains only even powers of the field these functions must be 
symmetric respect to $z=L/2$, namely $D(z)=D(L-z)$.

%%%%%%%%%%%%%%%%%%%%%%%%%%%%%%%%%%%%%%%%%%%%%%%%%%%%%%%%%%%%%%%%%%%%%%%%%%%%%%
\section{The scaling}
%%%%%%%%%%%%%%%%%%%%%%%%%%%%%%%%%%%%%%%%%%%%%%%%%%%%%%%%%%%%%%%%%%%%%%%%%%%%%%
\label{sec:scaling}

As we have seen we decided to work with a scaled field $\vp'(x)$, related to the variable 
$\vp(x)$, used for example in \cite{Fantoni21c}, by
\bq \label{eq:sp}
\vp(x) = a^{-3/2} \vp'(x).
\eq
In other words, we are renormalizing the bare field. This can be compared with the standard 
renormalization formula 
\bq
\vp(x) = Z^{1/2} \vp^{\rm ren}(x).
\eq
$\vp^{\rm ren}(x)$ is referred to as the renormalized field and $Z$ is called the renormalization 
constant. In this language, we are setting $Z = a^{-3}$.

At the same time, we are rescaling the coupling constant with
\bq \label{eq:sg}
g = a^3 g'.
\eq
In the Standard Model, the various coupling constants also need to be renormalized for the 
continuum limit to exist, but the renormalization is not simply given by a power of the 
lattice spacing. Instead, it needs to carefully be tuned to the cutoff and to the couplings. 
In perturbation theory of canonical $\vp^4$, the bare coupling constant can be expressed in 
terms of the renormalized one, order by order. The result consists of a series that starts 
with $g^{ren}$:
\bq
g = g^{\rm ren} + c_2 (g^{\rm ren})^2 + c_3 (g^{\rm ren})^3  + \ldots
\eq

The standard renormalization procedure is based on the fact that the Fourier transform of the 
renormalized two-point-function contains a pole at $p^2 = M^2$, where $M$ is the physical 
mass of the particle. The renormalization constant $Z$ is chosen such that the residue of 
this pole is equal to 1. This ensures, in particular, that $\vp^{\rm ren}(x)$ and $\vp(x)$ as 
well as $g^{\rm ren}$ and $g$ have the same dimension. Note that our rescaling (\ref{eq:sp}) and 
(\ref{eq:sg}) instead changes the dimension of these objects.

We will soon see, in our first case study below, that the expectation value 
$\langle\vp'(x)\rangle$ tends to a constant when $N$ becomes large. This means that 
the expectation value of the unscaled field, $\langle\vp(x)\rangle$, tends to infinity in
proportion to $N^{3/2}$ \cite{Fantoni21c,Fantoni21f}. 

As we are holding $g'$ constant, the unscaled coupling constant $g$ tends to zero in proportion to $1/N^3$. This suggests that, for the parameter values we consider, the connected Green’s functions of the unscaled model tend to those of a free scalar field.

%%%%%%%%%%%%%%%%%%%%%%%%%%%%%%%%%%%%%%%%%%%%%%%%%%%%%%%%%%%%%%%%%%%%%%%%%%%%%%
\section{Numerical results}
%%%%%%%%%%%%%%%%%%%%%%%%%%%%%%%%%%%%%%%%%%%%%%%%%%%%%%%%%%%%%%%%%%%%%%%%%%%%%%

Our PIMC simulations use the Metropolis algorithm \citep{Kalos-Whitlock,Metropolis} 
to calculate the ensemble average of Eq. (\ref{eq:EV}) which is a $N^n$ 
multidimensional integral. The simulation is started from the initial condition 
$\vp(x)=\epsilon>0$ for all lattice points $x$, with $\epsilon$ a small positive 
number. One PIMC step consisted in a random displacement of each one of the $N^n$ 
field values, $\vp(x)$, as follows
\bq \label{eq:move}
\vp\rightarrow\vp+(2\eta-1)\delta,
\eq
where $\eta$ is a uniform pseudo random number in $[0,1]$ and $\delta$ is the 
amplitude of the displacement. Each one of these $N^n$ moves is accepted if 
$\exp(-\Delta S)>\eta$ where $\Delta S$ is the change in the action due to the move  
(it can be efficiently calculated considering how the kinetic part and the 
potential part change by the displacement of a single $\vp(x)$)
and rejected otherwise. The amplitude $\delta$ is chosen in such a way to have 
acceptance ratios as close as possible to $1/2$ and is kept constant during the 
evolution of the simulation. One simulation consisted of 
$M$ PIMC steps. The statistical error on the average $\langle\calo\rangle$ will then 
depend on the correlation time necessary to decorrelate the property $\calo$, $\tau_
\calo$, and will be determined as $\sqrt{\tau_\calo\sigma_\calo^2/(MN^n)}$, where $
\sigma_\calo^2$ is the intrinsic variance for $\calo$. 

We used up to a lattice of $N^n=25^4=390625$ points ($N=25$) and up to 
$M=2\times 10^6$ corresponding to $MN^n$ PIMC displacement moves. 

\subsection{First case study}

In our simulation we first chose the following study case $m=g=L=\beta=1$ and 
$\epsilon=10^{-10}$.

Notice that the minima of the two symmetric potential wells in the semi-classical 
Hamiltonian density described by the function 
$f(\vp)=\half \vp^2+\vp^4+{\textstyle\frac{3}{8}}\vp^{-2}$, are at 
$\vp_{\pm}=\pm 2^{-1/2}\approx \pm 0.707107$. From our Monte Carlo simulations (see 
Teble \ref{tab:tabI}), it seems that the vacuum expectation value of the field 
(one-point-value), $V=\sum_x\langle\varphi(x)\rangle/N^n$, tends to these values 
in the continuum limit, $a=1/N\to 0$. Note that in some of our previous works 
\cite{Fantoni22a,Fantoni22c,Fantoni22d} where, instead of keeping the bare mass $m$
constant, we tuned it so to have a constant renormalized mass $m_R$ we found $V=0$ in all 
cases. This is due to the fact that as $N$ increases so does the necessary bare mass 
which keeps constant $m_R$. So that the two symmetric potential wells in the semi-classical
Hamiltonian density has minima that tends to zero and one experiences tunneling of the
potential barrier at $\vp=0$.

In Table \ref{tab:tabII} we show the values for the renormalized mass, $m_R$, and 
coupling constant $g_R$ at increasing values of $N=1/a$. We see that in
the continuum limit $\lim_{a\to 0}m_R=0$ and $\lim_{a\to 0}g_R=2$, meaning that 
$\lim_{a\to 0}\langle\tilde{\vp}(0)^4\rangle/\langle\tilde{\vp}(0)^2\rangle^2=1$

\begin{center}
\begin{table}[h!]
\caption{Renormalized mass $m_R$, renormalized coupling constant $g_R$, and one-point-value (vacuum expectation value of the field) $V=\sum_x\langle\varphi(x)\rangle/N^n$ for
$n=3+1$, $m=g=L=\beta=1$ and $N=L/a=4,7,13,25$. 
In our PIMC simulations we used Eqs. (\ref{eq:EV}) and (\ref{eq:scaled-affine-action}). }
\label{tab:tabI}
%{\scriptsize
\begin{center}
%\begin{ruledtabular}
\begin{tabular}{||c|c|c|c||}
\hline
\hline
$N$ & $m_R$ & $g_R$ & $V$\\ 
\hline
4          &  0.1421(2) & 2.01178(2)  &  0.7501(4)\\
7          &  0.0602(1) & 2.001129(4) &  0.7339(2)\\
13         &  0.0224(2) & 1.999894(5) &  0.7216(2)\\
25         &  0.0084(1) & 2.000048(4) &  0.7148(3)\\
\hline
\hline
\end{tabular}
\end{center}
%\end{ruledtabular}
%}
\end{table}
\end{center}

\begin{figure}[h!]
\begin{center}
\includegraphics[width=9cm]{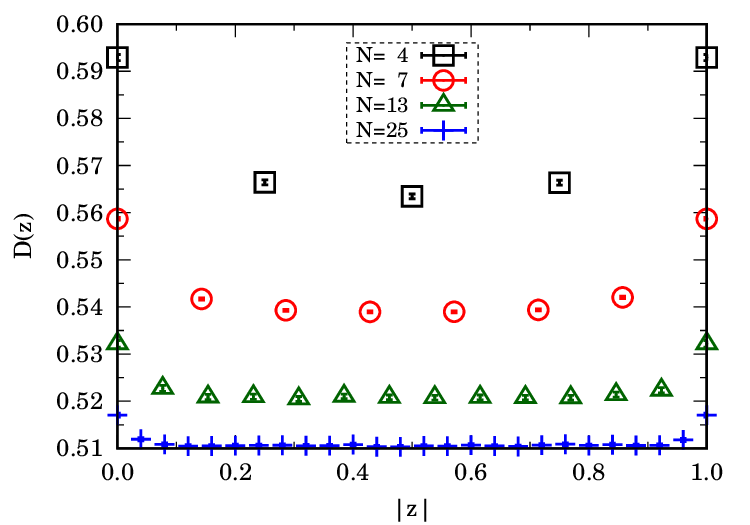}
\end{center}  
\caption{Two-point-function $D(z)=\sum_x\langle\varphi(x)\varphi(x+z)\rangle/N^n$ for
$n=3+1$, $m=g=L=\beta=1$ and $N=L/a=4,7,13,25$.
In our PIMC simulations we used Eqs. (\ref{eq:EV}) and (\ref{eq:scaled-affine-action}).}
\label{fig:green}
\end{figure}

In Figure \ref{fig:green} we show $D(z)$ at increasing values of $N=1/a$. From the plot 
of the simulation data we see that the function is symmetric respect to $z=1/2$ as 
expected, since the action only contains even powers of the field. 

\begin{figure}[h!]
\begin{center}
\includegraphics[width=9cm]{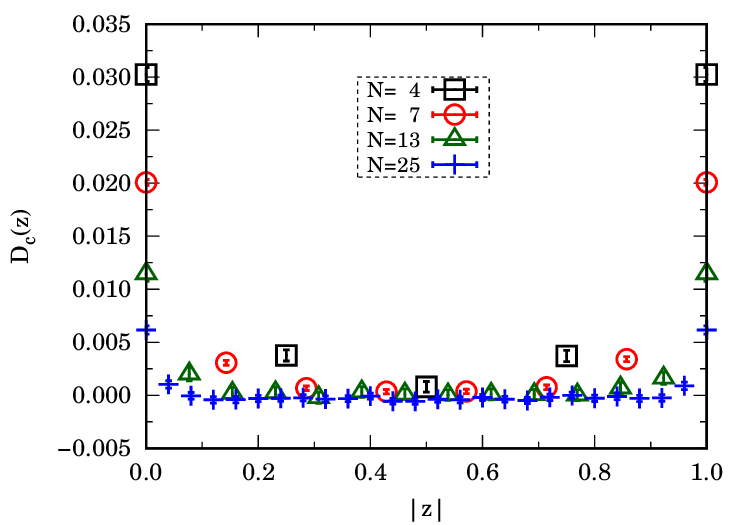}
\end{center}  
\caption{Two-point-connected-function 
$D_c(z)=D(z)-V^2$ for $n=3+1$, $m=g=L=\beta=1$ and $N=L/a=4,7,13,25$.
In our PIMC simulations we used Eqs. (\ref{eq:EV}) and (\ref{eq:scaled-affine-action}).}
\label{fig:greenc}
\end{figure}

In Figure \ref{fig:greenc} we show $D_c(z)$ at increasing values of $N=1/a$. From the 
plot of the simulation data we see that $\lim_{a\to 0}D(1/2)=0$. The width of the spike of 
$D_c(z)$ at $z=0$ seems to be related to the value of the renormalized mass $m_R$. 

Alternatively we could have adjusted, at each change of $N$, the value of the 
bare mass $m$ so to have a fixed value for the renormalized mass $m_R$. This would 
have resulted in a convergence towards a unique two-point-connected-function in
the continuum limit $N\to\infty$. We did not choose did strategy because it is numerically 
problematic to tune the bare mass so to have at each $N$ the same value for the 
renormalized mass. This was what we did in some of our previous papers 
\cite{Fantoni21c,Fantoni21e,Fantoni22a,Fantoni22b,Fantoni22c,Fantoni22d,Fantoni23b,
Fantoni23c,Fantoni23h}. \red{As explained in those works keeping the renormalized mass 
constant is extremely cumbersome due to the unavoidable sistematic numerical error that one
faces. It would have been then extremely difficult to obtain a reasonable comparison between 
the two point function at different $N$.}

\subsection{Second case study}

For the parameter values we just used, the box plays a crucial role: the bare Compton 
wavelength ($1/m$) is equal to the size $L$ of the box. In order for the box to be a purely 
technical device introduced to regularize the theory, it must be large compared to the 
correlation length of the model. At the same time, the lattice spacing must be small compared 
to it:
\bq
1/L \ll m \ll 1/a.
\eq

Therefore, next we considered the study case with $g=L=\beta=1$, $m=\sqrt{N}/L$
and $\epsilon=10^{-10}$, which should be much less affected by the presence of the box 
than the previous choice $m = 1/L$.

\begin{center}
\begin{table}[h!]
\caption{Renormalized mass $m_R$, renormalized coupling constant $g_R$, and one-point-value (vacuum expectation value of the field) $V=\sum_x\langle\varphi(x)\rangle/N^n$ for
$n=3+1$, $g=L=\beta=1$, $m=\sqrt{N}/L$ and $N=L/a=4,7,13,25$. 
In our PIMC simulations we used Eqs. (\ref{eq:EV}) and (\ref{eq:scaled-affine-action}). }
\label{tab:tabII}
%{\scriptsize
\begin{center}
%\begin{ruledtabular}
\begin{tabular}{||c|c|c|c||}
\hline
\hline
$N$ & $m_R$ & $g_R$ & $V$\\ 
\hline
4          &  0.1461(2) & 2.01199(2)  &  0.6672(4)\\
7          &  0.0627(1) & 2.001154(4) &  0.5992(3)\\
13         &  0.02463(5)& 1.999844(2) &  0.5169(2)\\
25         &  0.00867(5)& 2.000069(7) &  0.4359(3)\\
\hline
\hline
\end{tabular}
\end{center}
%\end{ruledtabular}
%}
\end{table}
\end{center}

Notice that the minima of the two symmetric potential wells in the semi-classical 
Hamiltonian density described by the function 
$f(\vp)=\half m^2\vp^2+\vp^4+{\textstyle\frac{3}{8}}\vp^{-2}$, $\vp_{\pm}(m)$, are such 
that 
\bq
\vp_\pm(m)=\pm 3^{1/4}(2m)^{-1/2}+{\rm O}(m^{-7/2}) ~~~\mbox{for $m\gg 1$}.
\eq
So with our choice of $m=\sqrt N$ we will find, in the continuum limit, 
$\lim_{N\to\infty}\langle\vp\rangle=\lim_{N\to\infty}\vp_+(\sqrt{N})=0$ in agreement 
with the results in Refs. \cite{Fantoni21c,Fantoni21e,Fantoni22a,Fantoni22c,
Fantoni22d,Fantoni23b,Fantoni23c,Fantoni23h}.

\begin{figure}[h!]
\begin{center}
\includegraphics[width=9cm]{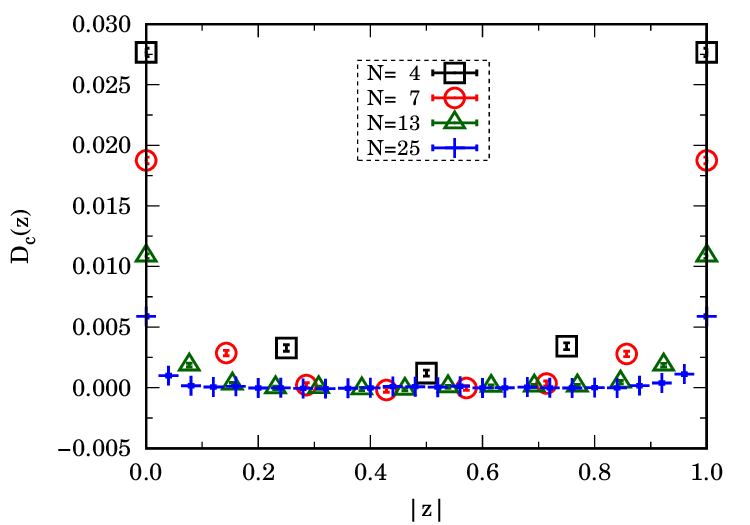}
\end{center}  
\caption{Two-point-connected-function 
$D_c(z)=D(z)-V^2$ for $n=3+1$, $g=L=\beta=1$, $m=\sqrt{N}/L$ and $N=L/a=4,7,13,25$.
In our PIMC simulations we used Eqs. (\ref{eq:EV}) and (\ref{eq:scaled-affine-action}).}
\label{fig:greenc2}
\end{figure}

From Fig. \ref{fig:greenc2} we see the continuum limit $N\to\infty$, of the scaled 
two-point-connected-function where, with an abuse of notation, we dropped from $D_c$, 
the prime, adopted rigorously in Section \ref{sec:scaling}. 
Respect to the work \cite{Fantoni21f} (see Fig. 3 there) which dealt with the unscaled 
free field case and $D_c(0)$ was found to increase with increasing $N$ we see how now 
the scaling has the effect of letting the value of $D_c'(0)$ decrease with increasing 
$N$, as shown in Fig. \ref{fig:greenc2}, since $\vp'=\vp/N^{3/2}$ and $D_c'=D_c/N^3$.
It is only tuning the bare mass $m$ so to have a constant renormalized mass $m_R$ for
each $N$, that we would find true convergence. Unfortunately this procedure is not 
easily accomplished numerically since for each $N$ we would have to make several 
test runs with different values of $m$ in order to find the value which keeps $m_R$
approximately constant. This procedure was nonetheless carried out in the following 
works \cite{Fantoni21c,Fantoni21e,Fantoni22a,Fantoni22c,
Fantoni22d,Fantoni23b,Fantoni23c,Fantoni23h}. 

%%%%%%%%%%%%%%%%%%%%%%%%%%%%%%%%%%%%%%%%%%%%%%%%%%%%%%%%%%%%%%%%%%%%%%%%%%%%%%
\section{Conclusions}
%%%%%%%%%%%%%%%%%%%%%%%%%%%%%%%%%%%%%%%%%%%%%%%%%%%%%%%%%%%%%%%%%%%%%%%%%%%%%%

In this paper, we represent $\pi(x)$ by $k(x)/\vp(x)$. To insure 
proper values for $\pi(x)$ it is necessary to restrict $0<\vp(x)<\infty$ as well 
as $0\leq|k(x)|<\infty$. Indeed such symbol change is able to treat Hamiltonian 
densities with an interaction $\vp(x)^4$. 
This leads to a completely satisfactory quantization of field theories 
using situations that involve scaled behavior leading to an unexpected, 
$\hbar^2/\vp(x)^2$ which arises only in the quantum aspects. Indeed, it is fair 
to claim that this symbol change leads to valid field theory quantizations.

Respect to the work \cite{Fantoni21f} which dealt with the free field case we 
here repeat that analysis but now for the $\vp^4$ interacting case.

We prove through path integral Monte Carlo computer experiments that the affine
quantization of the $\vp_4^4$ scaled Euclidean covariant relativistic field theory 
is a well defined quantum field theory with a well defined continuum limit of the 
one- and two-point-function, the Green's function.

The simple pseudo-potential $\propto\hbar^2/\vp^2$ stemming from the affine 
quantization procedure \cite{Klauder2020c} not only does not disturb the continuum 
limit, as we proved here, but in addition is able to render \red{non-trivial} the 
$\vp_4^4$ theory which is known \cite{Freedman1982} to be \red{trivial} when 
treated with the more commonly known \cite{Dirac} canonical quantization . 
 
\appendix
%%%%%%%%%%%%%%%%%%%%%%%%%%%%%%%%%%%%%%%%%%%%%%%%%%%%%%%%%%%%%%%%%%%%%%%%%%%%%%
\section{The extra ``$3/4$'' potential term}
%%%%%%%%%%%%%%%%%%%%%%%%%%%%%%%%%%%%%%%%%%%%%%%%%%%%%%%%%%%%%%%%%%%%%%%%%%%%%%
In order to explain the extra ``$3/4$'' potential term we use the fact that 
the operator corresponding to the affine field $\kappa$ will be the {\sl dilation} 
operator 
$\widehat{\kappa}=(\widehat{\pi}\widehat{\varphi}+\widehat{\varphi}\widehat{\pi})/2$ 
where the regularized basic quantum Schr\"odinger operators are given by 
$\widehat{\varphi}(x)=\varphi(x)$ and 
$\widehat{\pi}(x)=-i\hbar\delta_{\varphi(x)}=-i\hbar\delta/\delta\varphi(x)$ so that the 
commutator
$[\widehat{\varphi}(x),\widehat{\pi}(y)]=i\hbar\delta^s(x-y)$, where $\delta^s(x)$ is 
a $s$-dimensional Dirac delta function since 
$\delta_{\varphi(x)}\varphi(y)=\delta^s(x-y)$. Multiplying this by $\widehat{\varphi}$ 
we find 
$[\widehat{\varphi},\widehat{\varphi}\widehat{\pi}]=[\widehat{\varphi},\widehat{\pi}\widehat{\varphi}]=[\widehat{\varphi},\widehat{\kappa}]=i\hbar\delta^{s}\widehat{\varphi}$ 
which is only valid for $\varphi\neq 0$. Then
$\widehat{\kappa}=-i\hbar\{\delta_{\varphi(x)}[\varphi(x)]+\varphi(x)\delta_{\varphi(x)}\}/2=-i\hbar\{\delta^s(0)/2+\varphi(x)\delta_{\varphi(x)}\}$. 
Now, for $\varphi(x)\neq 0$, we will have that affine quantization sends 
$\widehat{\pi\mkern 0mu}^2(x)$ to
\bq \nonumber
\widehat{\kappa}(x)\varphi^{-2}(x)\widehat{\kappa}(x)&=&
-\hbar^2\{\delta^s(0)/2+\varphi(x)\delta_{\varphi(x)}\}\varphi^{-2}(x)
\{\delta^s(0)/2+\varphi(x)\delta_{\varphi(x)}\}\\ \nonumber
&=&\hbar^2(3/4)\delta^{2s}(0)\varphi^{-2}(x)-\hbar^2\delta_{\varphi(x)}^2\\
&=&\hbar^2(3/4)\delta^{2s}(0)\varphi^{-2}(x)+\widehat{\pi\mkern 0mu}^2(x).
\eq
%%%%%%%%%%%%%%%%%%%%%%%%%%%%%%%%%%%%%%%%%%%%%%%%%%%%%%%%%%%%%%%%%%%%%%%%%%%%%%
\bibliography{continuum}
%\bibliographystyle{prsty}

%%%%%%%%%%%%%%%%%%%%%%%%%%%%%%%%%%%%%%%%%%%%%%%%%%%%%%%%%%%%%%%%%%%%%%%%%%%%%%
%%%%%%%%%%%%%%%%%%%%%%%%%%%%%%%%%%%%%%%%%%%%%%%%%%%%%%%%%%%%%%%%%%%%%%%%%%%%%%
%%%%%%%%%%%%%%%%%%%%%%%%%%%%%%%%%%%%%%%%%%%%%%%%%%%%%%%%%%%%%%%%%%%%%%%%%%%%%%
\end{document}